\documentclass[12pt]{article}
\usepackage{epsf}
\usepackage{epsfig}
\usepackage{latexsym,amssymb,euscript}
\usepackage{amsfonts}
\usepackage{amsmath}

\begin{document}
\begin{center}

{\Large \bf MONOPOLES IN THE PLAQUETTE FORMULATION OF THE $3D$
$SU(2)$ LATTICE GAUGE THEORY}

\vspace*{0.6cm} {\bf O.~Borisenko$^*$\footnote{email:
oleg@bitp.kiev.ua}, \
S.~Voloshyn$^*$\footnote{email: billy.sunburn@gmail.com}} \\
\vspace*{0.5cm}
{\bf J. Boh\'a\v cik$^{\dag}$\footnote{email: Juraj.Bohacik@savba.sk}} \\
\vspace*{0.3cm} {\large \it $^*$N.N.Bogolyubov Institute for
Theoretical Physics, National Academy of Sciences of Ukraine, 03143
Kiev, Ukraine} \\ {\large \it $^{\dag}$Institute of Physics, Slovak
Academy  of Sciences, 84511 Bratislava, Slovakia}
\end{center}

\begin{abstract}
Using a plaquette formulation for lattice gauge models we describe
monopoles of the three dimensional $SU(2)$ theory which appear as
configurations in the complete axial gauge and violate the continuum
Bianchi identity. Furthemore we derive a dual formulation for the
Wilson loop in arbitrary representation and calculate the form of
the interaction between generated electric flux and monopoles in the
region of a weak coupling relevant for the continuum limit. The
effective theory which controls the interaction is of the
sine-Gordon type model. The string tension is calculated within the
semiclassical approximation.
\end{abstract}


\section{Introduction}

The problem of the permanent confinement of quarks inside hadrons
attracts attention of the theoretical physicists for the last three
decades (see \cite{greensite} and refs. therein for a review of the
problem). Two of the most popular and the most elaborated mechanisms
of confinement are based on the condensation of certain
topologically nontrivial configurations - the so-called center
vortices or monopoles. In this paper we are interested in the second
of these configurations. It was proposed in the context of continuum
compact three dimensional ($3D$) electrodynamics that the string
tension is nonvanishing in this theory at any positive coupling
constant  \cite{polyakov}. Configurations responsible for such
behaviour have been identified as monopoles of the compact theory.
The contribution of monopoles to the Wilson loop was estimated in
the semiclassical approximation. Later this consideration was
extended to $3D$ $U(1)$ lattice gauge theory (LGT) \cite{dualu1}. A
rigorous proof of the confinement was constructed in \cite{mack}.
While the monopoles of abelian gauge models can be given a gauge
invariant definition, it is not the case for nonabelian models (see,
however \cite{di_giacomo} for discussion of this point). The most
popular approach to the problem consists of a partial gauge fixing
such that some abelian subgroup of the full nonabelian group remains
unbroken. Then, one can define monopoles in a nonabelian theory as
monopoles of the unbroken abelian subgroup. Here we propose a
different way to identify monopole configurations in nonabelian
models. Its main feature is a complete gauge fixing. Monopoles
appear as defects of smooth gauge fields which violate the Bianchi
identity in the continuum limit, in the full analogy with abelian
models. Our principal approach is to rewrite the compact LGT in the
plaquette (continuum field-strength) representation and to find a
dual form of the nonabelian theory. The Bianchi identity appears in
such formulation as a condition on the admissible configurations of
the plaquette variables. This allows to reveal the relevant field
configurations contributing to the partition function and various
observables. Such a program was accomplished for the abelian LGT in
\cite{dualu1}. Here we are going to work out the corresponding
approach for nonabelian models on the example of $3D$ $SU(2)$ LGT.

Our strategy is to represent the  action of the model in the
plaquette formulation in the form that generalizes the abelian
"monopole + photon" effective action. In the  $SU(2)$  case the dual
gluon field carries colour index and the monopoles are coupled to
the length of the dual auxiliary field. As a result, the
gluon-monopole coupling is nonlinear. To treat  this problem we
solve classical equations with the help of a certain anzatz.
Performing standard procedure\cite{dualu1} we rewrite the effective
monopole action in the form of the sine-Gordon theory. In some
approximation we get an area law for the Wilson loop in the
fundamental (and all half-integer) representation. Results for
arbitrary representations of the Wilson loop are also discussed. An
approach similar in spirit to ours has been developed in \cite{Conr}
.

The paper is organized as follows. In section 2 we briefly review
the plaquette representation for gauge models. Section 3 is devoted
to an attempt to derive confinement in $SU(2)$ LGT. We begin with
discussing some problems of confinement in $3D$ $U(1)$ theory
including $j$-dependence of the string tension. In subsection 3.1 we
derive the effective monopole action. The derivation of the area law
is given in subsection 3.2. The results are summarized and discussed
in the section 4. In appendix A we give an approximate solution of
the classical equation of the sine-Gordon model for the adjoint
representation (also valid for $j=2$ abelian case). In appendix B we
discuss some aspects of the strong coupling expansion of the Wilson
loop in the plaquette representation.

\section{Plaquette formulation and monopoles}

The plaquette representation was invented originally in the
continuum theory by M.~Halpern \cite{plaqorig1} and extended to
lattice models by G.~Batrouni \cite{plaqorig2}. In this
representation the plaquette matrices play the role of the dynamical
degrees of freedom and satisfy certain constraints expressed by
Bianchi identities in every cube of the lattice. In papers
\cite{plaq}, \cite{plaq_full} we have developed a different
plaquette formulation which we outline below.

In the complete axial gauge
\begin{equation}
\label{func5} U_{3}(x,y,z)=U_{2}(x,y,0)=U_{1}(x,0,0)=I
\end{equation}
\noindent the partition function of $SU(N)$ LGT can be rewritten on
the dual lattice as \cite{plaq_full}
\begin{equation}
\label{dualpart} Z = \int \prod_{l}d V_{l} \exp \left[\beta \sum_{l}
\mathrm{Re \ Tr} V_{l} \right] \ \prod_{i=1}^4 \ \prod_{x(i)} \
J\left ( V_x^{(i)} \right ) \ .
\end{equation}
Here, $V_l$ is a plaquette (dual link $l$ in $3D$) matrix which
satisfies a constraint expressed through the group delta-function
\begin{equation}
\label{func2} J(V_x) = \sum_r d_r \chi_r \left ( V_x \right ) \ ,
\end{equation}
where the sum over $r$ is a sum over all representations of $SU(N)$
, $\chi_r$ is the character of the $r$-th representation and
$d_r=\chi_r(I)$. $V_x$ is a certain product of the  plaquette
matrices around a cube (dual site $x$) of the lattice taken with the
corresponding connectors. Connectors provide correct parallel
transport of opposite sites of a given cube for nonabelian theory.
In abelian models connectors are canceled out of the group
delta-functions. There appear four different types of connectors in
our construction. This distinction however is not important for our
purposes in this paper. Exact expressions for $V_x^{(i)}$ can be
found in \cite{plaq_full}.

We consider here the $SU(2)$ gauge group. We use the standard
parameterization of the group matrices through elements of an
algebra of $SU(2)$
\begin{equation}\label{V_l}
V_l = \exp \left[i \sigma_k \frac{\omega_k (l)}{2} \right] \ ,
\end{equation}
where $\sigma_k $ are  Pauli matrices. The constraint (\ref{func2})
expressed in terms of link angles $\omega_k (l) \equiv \omega_k
(x,n) $ on the dual lattice reads
\begin{equation}
\left[ \sum_k \frac{\omega_k^2 (x)}{2} \right]^{\frac{1}{2}} = 2 \pi
m (x) \ , \label{bianchi}
\end{equation}
where $m(x)$ are arbitrary integers. $\omega_k(x)$ can be expanded
into a power series
\begin{equation}
\omega_k (x) = \sum_{n=1}^3 \left ( \omega_k(x,n) -
\omega_k(x-e_n,n) \right ) + {\cal{O}}( \omega_k^2(l))  \ .
\end{equation}
Here, six links $l=(x,n)$ are attached to a site $x$ and
$\omega_k(l)$ are link variables dual to the original plaquettes. In
the continuum limit the last constraint reduces to the familiar
Bianchi identity if one takes $m(x)=0$ for all $x$. However, when
$m(x)$ differs from zero one gets a violation of the continuum
Bianchi identity at the point $x$. This is genuine feature of the
compact gauge models. Below we want to clarify a role of these
configurations in producing the string tension. Clearly, $m(x) \ne
0$ configuration corresponds to the monopole configuration of
nonabelian gauge field. Therefore, we may interpret the summation
over $m(x)$, appearing below, as a summation over monopole charges
that exist due to the periodicity of $SU(2)$ delta-function (in
close analogy with the $U(1)$ lattice model).

Substituting (\ref{V_l}) into (\ref{func2}) and making the Taylor
expansion of the original action around $\omega_k (l) \approx 0$ one
can prove that the partition function (\ref{dualpart})  can be
rewritten at $\beta \rightarrow \infty$ as \cite{2dsunlink}
\begin{gather}
Z = \int \prod_{l, k} d\omega_k(l)  \exp \left[ - \frac{\beta}{4}
\sum_{l,k}\omega_k^2 (l) \right] \prod_x \frac{|W_x|_{mod \ 2
\pi}}{\sin W_x}
\nonumber     \\
\prod_x \sum_{m(x)=-\infty}^{\infty}\int\prod_kd\alpha_k(x) \exp
\left [ -i\sum_k\alpha_k(x)\frac{\omega_k(x)}{2} + 2\pi im(x)\alpha
(x) \right ] \ , \label{PFwk}
\end{gather}
where $\alpha (x)=(\sum_k\alpha^2_k(x))^{1/2}$,
$W_x=\frac{1}{2}(\sum_k \omega_k^2 (x))^{1/2}$.

The formula (\ref{PFwk}) is the starting point in the construction
of an effective monopole theory. This representation, in fact,
generalizes the photon-monopole representation of the  $U(1)$ LGT to
the nonabelian model. Dual potentials  $\alpha_k(x)$ interact with
massless dual gluon fields $\omega_k(l)$ and with monopoles.  Unlike
the abelian case, the latter interaction is highly nonlinear.

The Wilson loop of the size $R\times T$ in some representation $j$
gets the form \cite{2dsunlink}
\begin{equation}
W_j(C) = \mathrm{Tr }_j \! \! \! \prod_{n=R/2-1}^{0} \! \left(
\prod_{z_1=0}^{z+T-1} \! V_1^{\dag}(x,y+2n+1,z_1) \! \! \!
\prod_{z_2=z+T-1}^0 \! V_1^{\dag}(x,y+2n,z_2) \right) \! .
\label{WLdual1}
\end{equation}
The  product runs over all dual links which belong to the minimal
surface bounded by the loop $C$. We have supposed, for simplicity,
that the loop contour lies in the $y-z$ plane, one side of the loop
lies in the plane $z=0$ and $R$, $T$ are even.

\section{Confinement in three dimensional $SU(2)$ LGT}

Let us remind some facts about $3D$ compact lattice electrodynamics.
As is known the original model can exactly be  rewritten in the form
of the  Coulomb gas of magnetic monopoles $m(x)$ interacting with an
electric current loop generated by sources
\begin{eqnarray}
S_{eff} = - \frac{1}{4\beta} h (b) G_{bb^{\prime}} h (b^{\prime})
-\pi^2\beta m (x) \ G_{x,x^{\prime}} \ m (x^{\prime})+i \pi h(b)
D_b(x^{\prime})m(x^{\prime})  \ , \label{U1_S_CL}
\end{eqnarray}
where $h(b)$
\begin{equation}
h (b) \ = \
\begin{cases}
j, \  b \in S^d \\
0, \  b \neq S^d \ ,
\end{cases}
\label{WL_sources}
\end{equation}
and sums over repeating indices are understood here and below. $S^d$
is a dual surface bounded by the Wilson loop $C$ in the
representation $j$. Here we have introduced the link Green functions
$G_{ll^{\prime}}$ and $D_l(x)$ (see their definitions and properties
in the Appendix of our paper \cite{plaq_full}). The first term in
(\ref{U1_S_CL}) corresponds to a perimeter contribution from dual
photons (confining logarithmic potential in $3D$).

Following the strategy of \cite{polyakov,dualu1} one can use the
dilute monopole gas approximation to perform summation over monopole
charges  $m (x)$. We skip all technical details which are well
known. The resulting theory appears to be of the sine-Gordon type
with exponentially small mass $m$. The semiclassical estimation of
the string tension for $j=1$ predicts $\sigma= \frac{1}{\pi^2
\beta}m$ which appears to be a lower bound on the exact string
tension\cite{mack}. Finding correct $j$ dependence of the string
tension remains an open problem. E.g., it has been argued
in\cite{polyakov1} that the  string tension has the following
dependence on $j$
\begin{equation}
\sigma_j =  \begin{cases} \sigma_{j=1} , \  j \text{ is  odd} \ ,\\
0 , \  j \text{ is  even}  . \end{cases} \label{strg_ten_pol}
\end{equation}

However, semiclassical approximation for the Wilson loops in higher
representations ($j\geq 2$) developed in \cite{Ambj_Greens} led to
the result
\begin{equation}\label{j_m}
\sigma \sim   j m \ .
\end{equation}
Moreover, as can be seen from bounds of \cite{mack}  (formula 8.3),
the string tension might well behave as
\begin{equation}\label{j^2_m}
\sigma \sim   j^2 m \ .
\end{equation}

Usually, semiclassical estimations are obtained in $1D$
approximation. It was stressed in \cite{Ambj_Greens}  that in this
approximation one cannot construct solutions leading to (\ref{j_m})
and/or (\ref{j^2_m}). One should probably go beyond $1D$ approach to
get  the correct $j$-dependence. A hint on this can be found in the
strong coupling expansion for the Wilson loop. Indeed, it is  a
planar diagram (minimal surface) contributing in the leading order
to the $j=1$ Wilson loop. When the size of the loop grows to
infinity, the  $1D$ approximation  can be justified.  However,
diagrams contributing to the $j>1$ Wilson loop a big size are
essentially three-dimentional. The $j^2$-dependence holds for the
loops of middle sizes (see \ref{abelianized_str_tens}). Therefore,
in both these cases the  $1D$ approximation is not valid. In
Appendix A we attempt to construct approximate solution to the full
$3D$ equation. Our result agrees with the formula (\ref{j_m}). All
this will be relevant in our discussion of $SU(2)$ model, namely
concerning solution of sine-Gordon equation in the  next subsection.

\subsection{Effective monopole model at large $\beta$}

Here we would like to calculate the contribution of monopole
configurations to the partition function and to the Wilson loop of
$SU(2)$ LGT. In doing so we use some approximations. First of them
is related to the fact that at large $\beta$ all plaquette matrices
fluctuate smoothly around unit matrix. But if we want to take into
account nontrivial monopole configurations one cannot expand fields
$\omega_k (l)$ and $\alpha_k (x)$ in (\ref{PFwk}) around the trivial
vacuum. We should make an expansion of the action, the invariant
measure and the Jacobian around nontrivial monopole configurations.
To construct such expansion we solve classical equations for fields
$\omega_k(l)$ and $\alpha_k (x)$ making use certain anzatz which
allows us to get rid of the nonlinear term $\sum_x m(x)\alpha (x)$.
After the solution is constructed we expand the effective action
around this solution. Actually, we restrict ourselves only to the
classical action. It should be mentioned that the connectors vanish
in this approximation due to the anzatz chosen. To take into account
their contribution, one has to keep fluctuations in the effective
action. In the present work we neglect connectors. The main
motivation for this approach comes from \cite{plaqorig2} where it
has been shown that connectors do not contribute to the fundamental
Wilson loop in the leading orders of the strong coupling expansion.
Thus, we may hope our approximation still captures the main
confining effect. Nevertheless, it turns out that connectors are
essential ingredients in reproducing the correct string tension for
the Wilson loop in higher representations, in particular in
reproducing the $N$-ality dependence, and this is true even in the
strong coupling regime. We demonstrate this in the Appendix B.

Consider the Wilson loop in the representation $j$. In the
parameterization (\ref{V_l}) the expectation value of $W_j(C)$
(\ref{WLdual1}) at $\beta \to \infty$ we present in the form
\begin{gather}
\left\langle W_j (C) \right\rangle \ = \ \frac{1}{2j+1} \
\left\langle \chi_j \left( \frac{\Omega_C}{2} \right) \right\rangle
\ =  \ \sum_{n=-j}^{j} \ \left\langle \cos n \Omega_C \right\rangle
\nonumber \\ = \ \frac{1}{2j+1}\ \sum_{n=-j}^j  \left. \left(1+
\frac{\partial}{\partial a} \right) \right|_{a=1} \int \prod_{k=1}^3
d \tau_k \frac{\delta \left(\sum_k \tau_k^2 - 1 \right)}{V(S^2)}
\left\langle e^{i a n \tau_k \Omega_k (C) } \right\rangle \ ,
\label{wilson_loop_su2}
\end{gather}
where
\begin{equation} \Omega_C \ = \ \left( \sum_k \Omega_k^2 (C) \right)^{\frac{1}{2}} \ ,
\ \Omega_k (C) = \sum_{l \in S^d} \omega_k(l) + {\cal{O}}(
\omega_k^2(l)) \ , \label{Omeg_C}
\end{equation}

Write down the classical equations for fields $\omega_k(l)$ and
$\alpha_k (x)$
\begin{gather}
\sum_{l \in x} \omega_k(l) + \epsilon^{kmn} \sum_{l < l^{\prime} \in
x} \omega_m (l) \omega_n (l^{\prime}) + ...  = 4
\pi m(x) \frac{\alpha_k (x)}{\alpha (x)} \ , \label{clas_eq_su2 1}  \\
- \frac{ \beta}{2} \omega_k (l) - i \frac{1}{2} \left(\alpha_k (x) -
\alpha_k (x + e_n)\right)  + ... =
 - i a n \tau_k h(l) \ , \label{clas_eq_su2 2}
\end{gather}
where $h(l)$ is defined in (\ref{WL_sources}). These equations are
too complicated to be solved in full. Since sources enter the
equation with a constant color vector $\tau_k$, we could look for
the solution in the form
\begin{equation}
\label{anzatz_tau} \omega_k (l)= \tau_k \omega^{s} (l) \ , \
\alpha_k (x) = \tau_k \alpha^{s} (x) \ .
\end{equation}
With this ansatz one gets the following equations for
$\omega^{s}(l)$ and $\alpha^{s}(x)$
\begin{eqnarray}
\frac{1}{2} \ \omega^{s} (x) = \frac{1}{2} \sum_n \left[\omega^{s}
(x, n)- \omega^{s} (x + e_n, n)\right ]
= 2 \pi m (x)  \ , \\
- \frac{2 \beta}{4} \omega^{s} (l) - i \frac{1}{2} \left[\alpha^{s}
(x) - \alpha^{s} (x + e_n)\right]  + i a n h(l) = 0 \ .
\end{eqnarray}
\noindent These equations can be easily solved as
\begin{gather}
\omega^s (l) = - \frac{an}{i \beta} G_{ll^{\prime}} h (l^{\prime}) +
2 \pi
D_l (y) m_y \ ,  \\
 \alpha^s (x) = G_{xx^{\prime}} \left( i 2 \pi \beta
m (x^{\prime}) + a n \sum_n \left[h(x^{\prime}, n)-h(x^{\prime}+
e_n, n) \right] \right) \ .
\end{gather}

Expanding now around the classical solutions and taking into account
that
\begin{equation} \frac{|W_x^{Cl}|_{mod \ 2 \pi}}{\sin W_x^{Cl}} = 1
\ , \label{W_sin W}
\end{equation}
the expectation value of $W(C)$ in (\ref{wilson_loop_su2}), is
presented in the form
\begin{gather}
\left\langle W_j (C) \right\rangle \ = \frac{1}{Z}  \ \frac{1}{2j+1}
 \ \sum_{n=-j}^j \left. \left(1+ \frac{\partial}{\partial a} \right) \right|_{a=1}
\sum_{ \{m(x)\} \ = \ - \infty}^{\infty} e^{S_{eff}} \ .
 \label{WL_clas}
\end{gather}
The effective action $S_{eff}$ reads
\begin{equation}
S_{eff} = - \frac{a^2 n^2}{2\beta} \sum_{b, b^{\prime} \in S^d}
G_{bb^{\prime}} - 2 \beta \pi^2 m(x)  G_{xx^{\prime}} m(x^{\prime})
+ 2 i a n \pi \sum_{b\in S^d} D_b(x^{\prime}) m(x^{\prime}) \ .
\label{Seffmon}
\end{equation}
This expression naturally generalizes the abelian analog
(\ref{U1_S_CL}) to the "monopoles + gluons" picture of $SU(2)$ case.
To perform the summation over monopole configurations $m_x=0, \pm 1$
we follow the strategy of Refs. \cite{dualu1,mack}. We omit all
technical details which are well known and present the result in the
form
\begin{gather}
\left\langle W_j (C) \right\rangle \ = \ \frac{1}{Z} \
\frac{1}{2j+1} \ \sum_{n=-j}^j \left. \left(1+
\frac{\partial}{\partial a} \right) \right|_{a=1} \nonumber \\
\times \ e^{- \frac{a^2 n^2}{2\beta} \sum\limits_{b, b^{\prime} \in
S^d} G_{bb^{\prime}}} \int_{-\infty}^{\infty} \prod_{x,k} d \phi_x \
e^{ - S_{SG} [\phi_x]} \ , \label{H_j}
\end{gather}
where $S_{SG}$ is the sine-Gordon action
\begin{equation}
 S_{SG}[\phi_x] =
\frac{1}{4 \beta}\sum_{x,n}( \phi_x - \phi_{x+n})^2 - \gamma \sum_x
 \cos [ 2 \pi (\phi_x +  a n \sum_{b\in S^d} D_b(x) ) ]
 \label{Seff3}
\end{equation}
with $ \gamma =  2 \exp [- 2 \pi^2 \beta G_0 (M) ] $.  Among other
properties the model (\ref{Seff3}) reveals the surface independence
of the Wilson loop. Namely, one can shift the surface $S^d$ without
any changes in the action (the properties of $D_b(x)$ guarantee
this).

Now, make a shift
\begin{equation}
\phi_x \rightarrow  -  a n  \sum_{b\in S} D_b(x) + \phi^c_x +\delta
\phi_x \ , \label{phi_shift}
\end{equation}
where $\phi_x^c$ is a classical solution of the saddle-point
equation and $\delta \phi_x$ is a fluctuation. Performing
perturbative in $\gamma$ integration over fluctuations and taking
derivatives in (\ref{H_j})   we get finally the following
representation for the Wilson loop
\begin{gather}
\left\langle \ W_j (C) \ \right\rangle \  =  \ \frac{1}{2j+1} \
\sum_{n=-j}^j \ e^{- \frac{ n^2}{2\beta} \sum\limits_{b,
b^{\prime} \in S^d} G_{bb^{\prime}}}  \nonumber  \\
\times \left[1 + \frac{2 n}{2 \beta} \sum_{y \in S^d} \left(
\phi^c_y -  \phi^c_{y+e_3} \right) - \frac{n^2}{\beta}
\sum_{b,b^{\prime} \in S^d} G_{bb^{\prime}} \right] \nonumber \\
\times \exp \left(- \frac{1}{4 \beta}\sum_{x,n}( \phi_x^c -
\phi^c_{x+n})^2 + \frac{2 n}{2 \beta} \sum_{y \in S^d} \left(
\phi^c_y - \phi^c_{y+e_3} \right) + \gamma \sum_x \cos [ 2 \pi
\phi_x^c ] \right) \ . \label{H_j_final}
\end{gather}

The result of perturbation theory can be easily recovered if one
takes  $\gamma=0$. Then $\phi^c_{x}\equiv0$ and we get
\begin{equation}\label{hkjt}
    \left\langle W_j (C) \right\rangle \  =   \
\frac{1}{2j+1} \ \sum_{n=-j}^j \ e^{- \frac{ n^2}{2\beta}
\sum\limits_{b, b^{\prime} \in S^d} G_{bb^{\prime}}}  \nonumber  \\
\left[1 - \frac{n^2}{\beta} \sum_{b, b^{\prime} \in S^d}
G_{bb^{\prime}} \right] \approx 1- \frac{j(j+1)}{2\beta}  2P(C) \ .
\end{equation}
For $R,T\to\infty$ one finds in $3D$
\begin{equation}
2P(C)=\sum_{l,l^{\prime}\in S^d} G_{ll^{\prime}} \approx
\frac{2}{\pi}(R\ln T + T\ln R) \ \ . \label{GLU1}
\end{equation}

In the next subsection we evaluate the monopole contribution to the
Wilson loop.

\subsection{The string tension in the semiclassical approximation}

To perform semiclassical calculations we take the continuum limit.
In this limit we get the following  saddle-point equation of the
sine-Gordon type
\begin{equation}
\label{sin-gordon_3D}
 \Delta \phi (x) =  4 \pi n \ \delta' (x) \ \theta (x;R,T) - m^2 \sin \phi (x)  \ ,
\end{equation}
where $\theta (x;R,T)$ is nonzero only if $x$ belongs to the surface
$S^d$. Here we have introduced the Debye mass
\begin{equation}
m^2 = 16 \pi^2 \beta e^{- 2 \pi^2 \beta G_0 (M)} \ .
\label{debyemass}
\end{equation}

Assuming that the Wilson loop is very large, we write down the
saddle-point equation (\ref{sin-gordon_3D}) as
\begin{equation}
\label{sin-gordon_1D} \phi^{''} (z) \ = \ 2 \pi n  \delta^{'} (z) -
m^2 \sin\phi (z) \ .
\end{equation}

Far from the boundaries of the contour $C$ the saddle-point equation
(\ref{sin-gordon_1D}) has the solution for $n=1/2$
\begin{equation}
\phi (z) \ = \
\begin{cases}
4 \arctan ( e^{- mz}), \ z > 0 \\
- 4 \arctan ( e^{ mz}), \ z < 0 \ . \label{polykov_sol}
\end{cases}
\end{equation}
This solution has an essential property
\begin{equation}
\phi (+0) - \phi (-0) \ = \ 2 \pi  \ .\label{cond_U1}
\end{equation}
From (\ref{polykov_sol}) one finds for the string tension
\begin{equation}\label{fund_str_ten}
    \sigma = \frac{1}{\pi^2 \beta} \ m \ .
\end{equation}

Unfortunately, there is no such simple solution of Eq.
(\ref{sin-gordon_1D}) for $n=1$ which represents adjoint Wilson
loop. We believe this is due to one dimensional approximation made
in going from (\ref{sin-gordon_3D}) to (\ref{sin-gordon_1D}), as we
have explained in the beginning of Section 3. Using idea from
\cite{Ambj_Greens} we have a free choice for the surface $S(C)$,
except for the requirement that $C$ is the boundary of $S$. In
particular, we could choose for the adjoint Wilson loop two sheets
that form two hemispheres with the loop $C$ being an equator (see
the Fig.(\ref{wils_loop_j=2-sphere})). For each sheet we now have a
discontinuity corresponding to (\ref{cond_U1}). In Appendix A we
describe the corresponding solution in more details. Our conclusion
is that the string tension for $n=1$ term in the expansion of the
Wilson loop will be twice the string tension of the fundamental
Wilson loop. In the general case we have
\begin{equation}\label{str_ten_n}
\sigma_{n} =   2 |n| \sigma_{1/2} \ .
\end{equation}
The solution (\ref{str_ten_n}) leads to the following result for the
Wilson loop
\begin{equation}\label{wil_loop_su2}
\left\langle W_{j}(C) \right\rangle \approx \  \frac{1}{2j+1} \
\sum_{n=-j}^j \ e^{ - 2 |n| \sigma_{1/2} A (C) - \frac{ 3
n^2}{2\beta} 2P(C)} \  ,
\end{equation}
where $A (C)$ is the area of the Wilson loop $C$. The second term in
the exponent is the leading term of the PT (see (\ref{GLU1})).
Finally, it is easy to obtain a general $j$-dependence for the
string tension
\begin{equation}\label{str_ten_su2}
\sigma =  \begin{cases} \sigma_{1/2} , \  j  \ \text{is  half-integer} \ , \\
0 , \  j \ \text{ is  integer} .\end{cases}
\end{equation}

\section{Summary and Discussion}

In this paper we calculated an effective model for the expectation
value of the Wilson loop in $3D$ $SU(2)$ LGT at large values of
$\beta$ . This model appears to be of a sine-Gordon type and could
be applied for all values of representations $j$ of $SU(2)$ group.
This model takes into account both the dual photons and the monopole
contributions. For all half-integer representations in the
semiclassical approximation we have found that the Wilson loop obeys
the area law. In the non-monopole sector of the model we recover the
result of the standard PT for the Wilson loop (perimeter law) after
integration over the fluctuations. In the abelian case our
calculations support the result of \cite{Ambj_Greens} that the
string tension of $U(1)$ LGT is proportional to $j$, i.e. $\sigma
\sim j$.

As is well-known, the string tension of the Wilson loops that are
non-trivial on the center obeys (\ref{string_tens_N-ality}) rather
than (\ref{abelianized_str_tens}) at the strong coupling, and it is
commonly accepted that (\ref{string_tens_N-ality}) represents true
asymptotic behaviour at large $\beta$ as well. Then,  the question
arises if the monopoles studied here can account for such behaviour.
First of all, the result for $j=1/2$ string tension is qualitatively
correct. It seems that our approach shows the expected $N$-ality
dependence of the string tension. But if we perform deeper analysis,
we can see that the formula (\ref{wil_loop_su2}) for the Wilson loop
do not correctly account for the perimeter law for integer
representations. The perimeter contribution is recovered in the PT
expansion at large $\beta$. Here, the question arises how to account
for the correct perimeter law decay of the Wilson loop. Both
connectors and higher order terms in the $1/\beta$ expansion are
presumably needed to achieve this. Though our representation
(\ref{wilson_loop_su2}) of the Wilson loop is exact, it is clear
that  the summation over $n$ together with approximations used makes
the whole theory equivalent to a set of $2j+1$ abelian-like theories
with  abelian $2n$-charge each. Also it is not clear how to get the
"Casimir scaling" behaviour in this approach. Our string tension has
$N$-ality dependence for all distances.  Nevertheless, \ monopole
contribution seems to be sufficient, if not necessary to get
confinement.

Finally, consider small Wilson loops. In this case to compute the
string tension it is allowed to expand the cosine function in the
effective model (\ref{H_j}). In the leading order and when $\beta
\to \infty$ and the size of the loop $C$ is fixed one obtains
\begin{gather}\label{casimir_scal}
    \left\langle W_j (C) \right\rangle \  =   \
\frac{1}{2j+1} \ \sum_{n=-j}^j \ e^{- \frac{ n^2}{2\beta} \sum_{b,
b^{\prime} \in S} G_{bb^{\prime}}(m)} \left[1 - \frac{n^2}{\beta}
\sum_{b, b^{\prime} \in S} G_{bb^{\prime}} (m) \right] \nonumber \\
\approx 1- \frac{j(j+1)}{2\beta} \left( 2P(C) + \sigma_{1/2}A (S)
 \right) \approx e^{- \frac{j(j+1)}{2\beta} \left( 2P(C) + \sigma_{1/2}A (C) \right)
 } \ ,
\end{gather}
where  $G_{bb^{\prime}} (m)$ is massive link Green function. This is
nothing but expected Casimir scaling of the string tension.

Reconstruction of the true string tension dependence would require
more refined analysis. The hint on this comes from the strong
coupling expansion in the plaquette formulation. Connectors of the
Bianchi identities do not contribute to the fundamental string
tension in the lowest order of small $\beta$-expansion. However,
connectors appear to be necessary to get the correct result
(\ref{string_tens_N-ality}) for all higher representations (see
Appendix B). In our derivation of the effective model we had no
choice but to neglect contribution from connectors to make the
problem solvable. Had we been able to include connectors in our
model we would probably have recovered the correct dependence. Such
a possibility is currently under investigation.

An approach similar in spirit to ours was developed in \cite{Conr} .
We believe, however that our effective model is more trustful. In
particular, we think the expectation value of the Wilson loop in
\cite{Conr} depends on the shape of the surface $S$ bounded by the
loop $C$. This is obviously unphysical property which our model is
free of.


\appendix

\section{Solution of the sine-Gordon equation for $j=1$ case}

Consider original  $3D$  sine-Gordon equation (\ref{sin-gordon_3D})
for the case $j=1$. To solve  (\ref{sin-gordon_3D}) we use an anzatz
of a general type
\begin{equation}
    \phi (x) =  4 \arctan  \omega  \ ,
\label{equatsadpoint1}
\end{equation}
where $\omega \equiv \omega (x)$ obeys the equation
\begin{equation}
    \triangle \omega - \frac{2 \omega \left(\nabla \omega \right)^2}{1+\omega^2}
    = - m^2 \frac{\omega (1-\omega^2) }{1+\omega^2} \ .
\label{omega_eq}
\end{equation}
We have been able to solve equation (\ref{omega_eq}) in the limiting
cases $|\omega| \ll 1$ or $|\omega| \gg  1$. This seems to be
sufficient to construct the solution with desired properties.
\begin{figure}[t]
\centerline{\epsfxsize=6cm \epsfbox{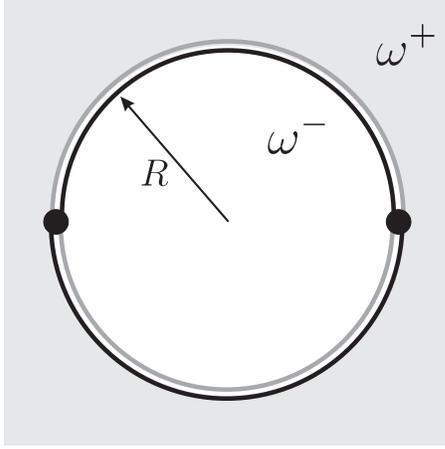}}
\caption{\label{wils_loop_j=2-sphere} Cross-section of the Wilson
loop surface deformed to two hemispheres of radius $R$.}
\end{figure}
For $|\omega| \ll 1$ one finds
\begin{equation}
   \triangle \omega = - m^2 \omega  \ ,
\label{omega_eq_case_ll_1}
\end{equation}
while for $|\omega| \gg 1$  (\ref{omega_eq}) becomes
\begin{equation}
    \frac{\triangle \omega }{\omega} - \frac{2 \left(\nabla \omega
    \right)^2}{\omega^2} = + m^2  \ . \label{omega_eq_case_gg_1}
\end{equation}
If $\omega$ is a solution of (\ref{omega_eq_case_ll_1}), then
$1/\omega$ is solution of (\ref{omega_eq_case_gg_1}).

Let us take for simplicity the circular Wilson loop. Then, the
surface $S$ is deformed into two hemispheres of the radius $R$
attached to the loop $C$ (i.e., the loop becomes an equator). Let
$\omega^+ (r,\theta)$ ($\omega^- (r,\theta)$) be solution of
Eq.(\ref{omega_eq_case_ll_1}) corresponding to outer (inner) regions
as shown in Fig. (\ref{wils_loop_j=2-sphere}). In this case the
condition (\ref{cond_U1}) reads
\begin{equation}\label{cond_sphere_1}
    \omega^+ (R,\theta) \ \omega^-(R,\theta) = 1 \ .
\end{equation}

The solution of eq. (\ref{omega_eq_case_ll_1}) can be chosen in the
form
\begin{equation}\label{jytkjm}
    \omega^+ = A \ (-\frac{m}{r}- \frac{1}{r^2}) e^{-mr} \cos\theta \ .
\end{equation}
This solution is easiest one that have an appropriate dependence
from $\theta$ and nothing but a spherical analog of the behaviour
$e^{-mz}$ valid in 1D case. It is valid for the outer region $r > R$
(see the Fig. (\ref{wils_loop_j=2-sphere})).

The constant $A$ can be fixed from the condition that at large $R
\gg 1/m$ the surface locally looks like plane and, therefore the
solution $\omega^+$ approaches exact $1D$ solution $e^{-mz}$. Thus,
$A = - \frac{R}{m}$. It is clear that our spherical solution should
coincide with $1D$ one for all values of $\theta$, so we believe the
correct $\theta$ dependence is
\begin{equation}\label{htrjyt}
    { \rm sign} (z) \simeq \frac{z}{r} + ... \ ,
\end{equation}
which reduces to $z / r = \cos \theta $ only for small $\theta$.
Hence, the asymptotic of the exact solution reads
\begin{equation}\label{om+}
    \omega^+ = \frac{R}{r} \ \left(1  + O \left(\frac{1}{mr} \right) \right) \ e^{-mr}
    \ { \rm sign} (z) \ .
\end{equation}

To meet the condition (\ref{cond_sphere_1}) we take for the inner
region ($r < R$) the function $\omega^-=1/\omega^+$ which is a
solution  of (\ref{omega_eq_case_gg_1}). Combining solutions for two
regions one gets
\begin{equation}
\omega \ = \
\begin{cases}
\frac{R}{r} e^{-mr} \ { \rm sign} (z)  ,  \ r > R  \\
\frac{r}{R} e^{mr} \ { \rm sign} (z)  ,  \ 0 < r < R   .
\label{full_sol}
\end{cases}
\end{equation}
Substituting the solution (\ref{full_sol}) into the action
\begin{equation}\label{nni}
    S (\omega) =  - \frac{1}{ \pi^2 \beta} \int d^3 x  \frac{(\nabla \omega)^2 + m^2
\omega^2}{(1 + \omega^2)^2}
\end{equation}
we finally obtain in agreement with \cite{Ambj_Greens}
\begin{equation}\label{adj_str_ten}
    \sigma \simeq 2 \ \frac{ m }{\pi^2 \beta  } \ .
\end{equation}

\section{Strong coupling expansion of Wilson Loop in plaquette representation}

\begin{figure}[t]
\centerline{\psfig{file=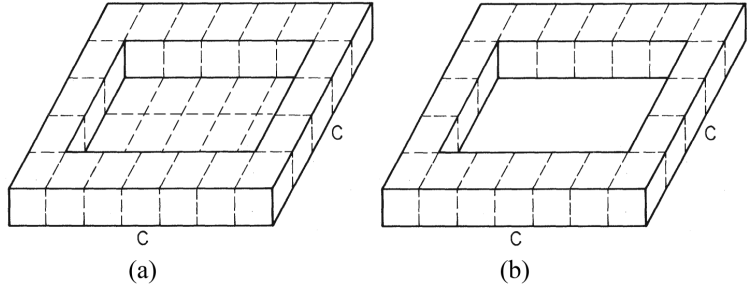,width=10cm}}
\caption{\label{strong_coupl_diagram-2} Strong-coupling diagrams
that represent $N$-ality  dependence: (a) area law  and (b)
perimeter law.}
\end{figure}

In this appendix we remind briefly the behaviour of the string
tension in different representations in the strong coupling region.
Then, we discuss how this behaviour can be understood from the point
of view of the plaquette formulation. For $U(1)$ gauge group the
string tension $\sigma_j$ in the representation $j$ behaves as
\begin{equation}
\label{abelianized_str_tens} \sigma_j \ = \
\begin{cases}
j \sigma_{j=1}, \ \mbox {big  \ Wilson \ loop} \ , \\
C_j \sigma_{j=1}, \ \mbox{middle \ Wilson \ loop} \ .
\end{cases}
\end{equation}
where $C_j=j^2$ is the quadratic Casimir operator. The diagram of
the "sandwich" type is responsible for the first behaviour, while
the second contribution is due to a planar diagram which covers the
Wilson loop in the representation $j$. Both types of diagrams exist
in $SU(N)$ models as well and contribute to the expectation value of
the Wilson loop. The corresponding string tension behaves like in
(\ref{abelianized_str_tens}), where one should take $\sigma_f$ in
the fundamental representation instead of $\sigma_1$. However,
different types of diagrams define the behaviour of asymptotically
large Wilson loops in non-abelian models. For all representations
$j$ which are trivial on the center $Z(N)$ there exists diagram of
type "tube" that leads to the perimeter law fall-off of the Wilson
loop. For all representations $j$ which transform non-trivially
under $Z(N)$ one has a combination of "tube" and planar diagram,
where plaquettes in the minimal surface are taken in the fundamental
representation, see Fig.\ref{strong_coupl_diagram-2}. Thus, the
string tension depends crucially on the $N$-ality of the
representation $j$ and equals
\begin{equation}
\label{string_tens_N-ality} \sigma_j \ \sim \
\begin{cases}
\sigma_{f}, \  \mbox{nonzero} \ N\mbox{-ality} \ , \\
0, \  \mbox{zero} \  N \mbox{-ality} \ .
\end{cases}
\end{equation}

The strong coupling expansion for nonabelian theories is an
expansion toward restoration of the Bianchi identity
\cite{plaqorig2}. For example, the leading term in the expansion of
the fundamental Wilson loop does not include contribution from the
connectors. Nevertheless, the connectors play crucial role as
building blocks of strong-coupling diagram that represents $N$-ality
dependence (\ref{string_tens_N-ality}). To see this, wright down the
expression of the Wilson loop (\ref{WLdual1}) in some representation
$j$ using the plaquette formulation
\begin{gather}
\label{wilson_loop_expan} \left\langle W_j (C) \right\rangle  =
Z^{-1} \ \int \prod_{l}d V_{l}  \ \prod_{i=1}^4 \ \prod_{x(i)} \
\sum_{\lambda_x = 0, 1/2, 1, ...}
d_{\lambda_x} \chi_{\lambda_x} \left(V_x^{(i)} [V_{l}] \right)  \nonumber  \\
\times \prod_l \left  [ 1 + \sum_{r \neq 0} d_r a_r (\beta) \chi_r
(V_{l}) \right ] \ \chi_{j} \left(\prod_{l \in S_d} V_l \right) \ .
\end{gather}
At small $\beta$ coefficients $a_r(\beta)$ are given by, e.g. for
$SU(2)$
\begin{equation}\label{a_r_asymptotics}
a_r (\beta) = \frac{I_{2r+1} (\beta)}{I_1 (\beta)} =
\frac{\beta^{2j}}{(2j+ 1)!} + O (\beta^{2j+2}) \ .
\end{equation}
For both integer and half-integer $j$ there exist contributions of
the form
\begin{equation}
 \left [
\frac{I_{2j+1}(\beta)}{I_1(\beta)} \right ]^{A(C)} \left [ 1+2A (C)
\left ( \frac{I_{2j+1}(\beta)}{I_1(\beta)} \right )^4 \right ] \ ,
\label{wilson-su_2_d2}
\end{equation}
which can be associated with the Casimir scaling. The leading term
in the above contribution comes from the action and neglects the
Bianchi constraint completely. The first correction includes one
term from Bianchi constraint, namely term $\lambda_x=j$. However,
the contribution from connectors is trivial (i.e., during the
invariant integration over plaquettes, the plaquettes from
connectors compensate themselves as a given plaquette appears twice
in a connector).

Main building block of three-dimensional diagram is Bianchi identity
that lives on the cube. To reproduce the "sandwich" type diagram one
has to put $\lambda_x=0$ for all cubes except those which have a
plaquette in common with the minimal surface bounded by the loop.
For such cubes one should take $\lambda=j-1/2$. In the next plane we
take $\lambda=j-1$ and so on. Combining this with fundamental
plaquettes from the action results in the contribution of the form
\begin{equation}
\left [ \frac{I_{2}(\beta)}{I_1(\beta)} \right ]^{j A(C)} e^{ -
\widetilde{\mu}_j P(C) } + ... \ . \label{wilson-su_2_d3-sand}
\end{equation}
At the invariant integration connectors are dropped out. It is easy
to see the diagrams of this type could be constructed without
connectors at all (i.e., with abelianized Bianchi identity) at least
in the lowest order.  So, in this case the contribution from
connectors is trivial.

However, one cannot get the strong coupling diagram that represents
$N$-ality dependence of the string tension without connectors. As
example, consider the $j=1$ case. First, we construct "tube" from
cubes in the fundamental representation (see
Fig.(\ref{strong_coupl_diagram-2})) and cover all outer plaquettes
by plaquettes taken from the action. Second, as is seen from formula
(\ref{WLdual1}) to get a nontrivial result of the invariant
integration we need to cover all plaquettes from the surface by some
plaquettes from additional Bianchi identities in the representation
$j=1$. Third, if we omit connectors from all Bianchi identities we
shall get a zero result due to integration over outer noncompensated
plaquettes from these additional cubes. Clearly, it is impossible to
build such "tube" from  plaquettes of abelianized Bianchi identity.
A result of the invariant integration would give a vanishing
contribution in this case. Thus, connectors appear to be a necessary
element in constructing correct $N$-ality dependence.

\section*{Acknowledgments}

Authours thank M.~Polykarpov and \v S.~Olejn\' ik for stimulating
discussions. This work was supported by the grant "Vacuum structure
and confinement mechanism in SU(N) gauge theories" between Slovak
and Ukrainian Academy of Sciences.


\end{document}